\documentclass[aps,prl,superscriptaddress,twocolumn,showpacs,amsmath,amssymb]{revtex4}
\usepackage{graphicx}% Include figure files
\usepackage{dcolumn}% Align table columns on decimal point
\usepackage{bm}% bold math

\begin{document}
%\preprint{APS/123-QED}

\title{Coexistence of amplitude and frequency modulations in intracellular calcium dynamics}% Force line breaks with \\

\author{Maurizio \surname{De Pitt\` a}}
\affiliation{School of Physics \& Astronomy, Tel-Aviv University,
69978
Tel-Aviv, Israel}%
\affiliation{Interdepartmental Research Center ``E. Piaggio", University of Pisa, 56125 Pisa, Italy}
\author{Vladislav Volman}
\affiliation{Center for Theoretical Biological Physics, University of California at San Diego, La Jolla, CA 92093-0319, USA}%
\affiliation{Computational Neurobiology Laboratory, The Salk Institute for Biological Studies, La Jolla, CA 92037, USA}%
\author{Herbert Levine}
\affiliation{Center for Theoretical Biological Physics, University of California at San Diego, La Jolla, CA 92093-0319, USA}%
\author{Giovanni Pioggia}
\affiliation{Interdepartmental Research Center ``E. Piaggio", University of Pisa, 56125 Pisa, Italy}
\author{Danilo \surname{De Rossi}}
\affiliation{Interdepartmental Research Center ``E. Piaggio", University of Pisa, 56125 Pisa, Italy}
\date{\today}
\author{Eshel Ben-Jacob}%
\email[Corresponding author: ]{eshel@tamar.tau.ac.il}%
\affiliation{School of Physics \& Astronomy, Tel-Aviv University, 69978 Tel-Aviv, Israel}%
\affiliation{Center for Theoretical Biological Physics, University of California at San Diego, La Jolla, CA 92093-0319, USA}%

\begin{abstract}
The complex dynamics of intracellular calcium regulates cellular responses to information encoded in extracellular signals. Here, we study the encoding of these external signals in the context of the Li-Rinzel model. We show that by control of biophysical parameters the information can be encoded in amplitude modulation, frequency modulation or mixed (AM and FM) modulation. We briefly discuss the possible implications of this new role of information encoding for astrocytes.
\end{abstract}

\pacs{Valid PACS appear here}%

\keywords{astrocyte, calcium, excitability, information, Bautin, cusp}%

\maketitle

Many cells use calcium signaling to carry information from the extracellular side of the plasma membrane to targets in their interior \citep{BerridgeBootman2000}.~This information serves many different purposes, from triggering the developmental program of fertilized mammalian eggs to mediating neural activity and learning or inducing cell death. Since virtually all cells employ a network of biochemical reactions for Ca$^{2+}$ signalling, much effort has been devoted to understand the functional role of the Ca$^{2+}$ response and to decipher how this complex dynamical response is regulated by the biochemical network of signal transduction pathways \citep{PolitiHofer2006,KummerBaier2000}.~About a decade ago, several experiments indicated that Ca$^{2+}$ signals in response to external stimulus can encode  information via frequency modulation (FM) or in some other cases via amplitude modulation (AM) \citep{Berridge1997}.~Consequently, it has been shown that these observations can be captured separately by minimal models consisting of two dynamical variables such as the Li-Rinzel \citep{LiRinzel1994} or the Dupont-Goldbeter models \citep{DupontGoldbeter1993}.~It was also shown that higher-order models (ones with several dynamical variables and/or intracellular diffusion mechanisms) can exhibit different and more advanced encoding modes \citep{FalckeRev2004}.\\
\indent
Here we propose that under certain conditions, heterogeneous dynamics of intracellular Ca$^{2+}$ could also be explained by opportune parameter modulation within minimal models. More specifically, we employ arguments of bifurcation theory to illustrate that within the minimal Li-Rinzel model the same cell could encode the information about external stimuli in amplitude modulation (AM) of calcium oscillations, in frequency modulation (FM) or in both (AFM). Our work is motivated by the calcium signalling in astrocytes, a predominant non-neuronal (glial) cell type that plays a crucial role in the regulation of neuronal activity \cite{NadkarniJung2003,VolmanNeurComp2007}. We explain why for this case, our results can be crucial for a better understanding of synaptic information transfer and propose that they might be equally important for better understanding of other examples of processes regulated by Ca$^{2+}$ signalling.
\begin{table}[tp!]
\centering{
\begin{tabular}{|c|c|c|c|}
\hline
$v_{1}$ & $6 sec^{-1}$ & $d_{1}$ & $0.13 \mu M$ \\
\hline
$v_{2}$ & $0.11 sec^{-1}$ & $d_{2}$ & $1.049 \mu M$  \\
\hline
$v_{3}$ & $0.9 \mu M sec^{-1}$ & $d_{3}$ & $0.9434 \mu M$  \\
\hline
$C_{0}$ & $2 \mu M$ & $d_{5}$ & $0.08234 \mu M$ \\
\hline
$c_{1}$ & $0.185$ & $a_{2}$ & $0.2 \mu M^{-1}sec^{-1}$ \\
\hline
$K_{3}$ & $0.1 \mu M$ & $$ & $$ \\
\hline
\end{tabular}
} \caption[]{\footnotesize Parameters used in the original Li-Rinzel model.}
\label{table-table1}
\end{table}

Calcium dynamics is controlled by the interplay of calcium-induced
calcium release, a nonlinear amplification process regulated by
the calcium-dependent opening of channels to Ca$^{2+}$ stores such
as the endoplasmic reticulum (ER), and by the action of active
transporters (SERCA pumps) which enable a reverse flux. The
dynamical variables of LR model, that is studied here, are the
free cytosolic Ca$^{2+}$ concentration $(C)$, and the fraction of
open inositol trisphosphate (IP$_{3}$) receptor subunits, $h$:
\begin{eqnarray}
\dot{C}&=&J_{chan}(C,I)+J_{leak}(C)-J_{pump}(C)\\
\dot{h}&=&\frac{h_{\infty}-h}{\tau_{h}}
\end{eqnarray}
The dynamics of $C$ is controlled by three fluxes, corresponding
to: 1.~a passive leak of Ca$^{2+}$ from the ER to the cytosol,
$(J_{leak})$; 2.~an active uptake of Ca$^{2+}$ into ER,
$J_{pump}$, due to the action of the pumps; 3.~a Ca$^{2+}$ release
$(J_{chan})$ that is mutually gated by Ca$^{2+}$ and the inositol
trisphosphate (IP$_{3}$) concentration, $(I)$:
\begin{subequations}
\label{eqs:Fluxes}
\begin{eqnarray}
J_{leak}(C) &=& v_{2}\left(C_{0}-\left(1+c_{1}\right)C\right)\\
J_{pump}(C) &=& \frac{v_{3}C^{2}}{K_{3}^{2}+C^{2}}\\
J_{chan}(C,I) &=& v_{1}m_{\infty}^{3}h^{3}\left(C_{0}-\left(1+c_{1}\right)C\right)
\end{eqnarray}
\end{subequations}
The gating variables and their time-scales are given~by:
\begin{align*}
m_{\infty} & = \frac{I}{I+d_{1}}\frac{C}{C+d_{5}} & h_{\infty} & = \frac{Q_{2}}{Q_{2}+C} & \tau_{h} & = \frac{1}{a_{2}(Q_{2}+C)}
\end{align*}
with  $Q_{2} = d_{2}\frac{I+d_{1}}{I+d_{3}}$. The level of IP$_{3}$ is directly controlled by signals impinging on the cell from its external environment. In turn, the level of IP$_{3}$ determines the dynamical behavior of the above model. One can therefore think of the Ca$^{2+}$ signal as being an encoding of information regarding the level of IP$_{3}$.

\begin{figure}[tb!]
\includegraphics[width=0.4\textwidth]{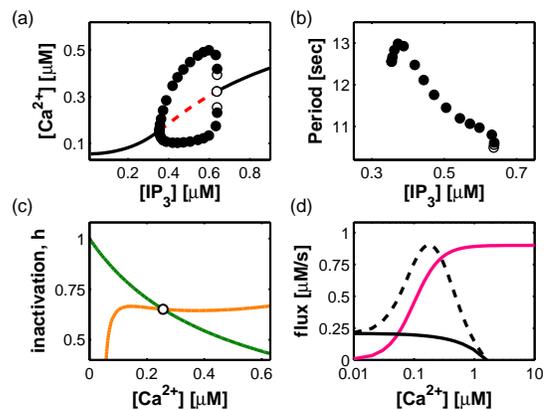}
\caption{\label{fig1:OrLR} \textbf{The Li-Rinzel model}.
\textit{(a)}~Bifurcation diagram for the original set of parameters of the Li-Rinzel model: ($-$)~stable fixed points, ($\cdots$)~unstable ones, ($\bullet$) stable limit cycles, ($\circ$)~unstable ones. Oscillations are born via supercritical Hopf bifurcation at [IP$_{3}$]$\simeq 0.355$ $\mu$M and die via subcritical Hopf bifurcation at [IP$_{3}$]$\simeq 0.637$ $\mu$M. While the amplitude changes, the frequency is nearly constant \textit{(b)}. \textit{(c)}~Nullclines (green: $h$, orange: $C$) for the case of an unstable point. \textit{(d)}~At basal IP$_{3}$ levels ($\simeq 0.015$ $\mu$M) $J_{pump}$ (red curve) intersects $J_{rel}$ (\textbf{$-$}) at a calcium-level such that ${J}'_{rel}<0$. This situation also occurs at  higher IP$_{3}$ when $J_{rel}$ becomes bell-shaped.}
\end{figure}

%%Carmignoto00
The original set of biophysical parameters, as given in table~\ref{table-table1}, corresponds to AM encoding. For these parameters, the phase-plane and bifurcation analysis reveals that, at $I\approx0.355\mu M$, limit-cycle Ca$^{2+}$ oscillations emerge through a supercritical Hopf bifurcation.~From fig.~\ref{fig1:OrLR}a it is evident that the amplitude of Ca$^{2+}$ oscillations increases between the two bifurcation points, while fig.~\ref{fig1:OrLR}b shows that the frequency of the oscillations is almost constant~-~hence the term ``Amplitude Modulation".~Amplitude modulation by IP$_{3}$ has been observed in many experiments; however new findings indicate that, under some conditions, variations (by external stimulation) in the level of IP$_{3}$ can also lead to frequency modulation \cite{Pasti1997}. These observations motivated us to re-examine the LR model to investigate if changes in the biophysical parameters could lead to frequency-modulated dynamics. A nonlinear system can exhibit frequency modulation in the presence of saddle-node bifurcation \cite{Izhikevich2000}. This latter describes a transition of a system from an excitable state (in which there are three fixed points: stable, unstable and a saddle) to a limit cycle. At a certain value of the control parameter $I=I_{sn}$, the stable and saddle fixed points coalesce and the only remaining attractor is a limit cycle. The frequency of the oscillations of the limit cycle can be very sensitive to the distance from the bifurcation point (i.e.: $I-I_{sn}$), whereas the amplitude remains almost constant~\cite{RinzelErmentrout1989}.

We have explored the range of parameters for which the LR system can exhibit a saddle-node bifurcation with the level of IP$_{3}$ being the control parameter. We found that K$_{3}$ (the affinity of the active SERCA pump), $d_{5}$ (the receptor affinity for IP$_{3}$) and $v_{2}$ (the rate of Ca$^{2+}$ leakage from the ER) all can regulate the switching between AM and FM encoding dynamics. We further discovered the existence of a new dynamical regime in which the variations of the IP$_{3}$ are co-encoded both in amplitude and frequency modulations (``AF Modulations"). This AFM dynamics exists for higher levels of the cell-averaged resting Ca$^{2+}$ concentration C$_{0}$ as well as for lower $v_{3}$ rates of Ca$^{2+}$ uptake by SERCAs. We present a biophysical picture of these different regimes and comment on the physiological implications of these results with particular attention to astrocytes. Although there have been some earlier indications that the LR model can encompass excitable behavior \cite{Slepchenko2000,StamatakisMantzaris2006}, these works did not present a complete analysis nor a biophysical picture. For brevity we consider only the case of varying $K_{3}$.\\
%%
%%
%%
%% Figure 2
\begin{figure}[htb!]
\includegraphics[width=0.4\textwidth]{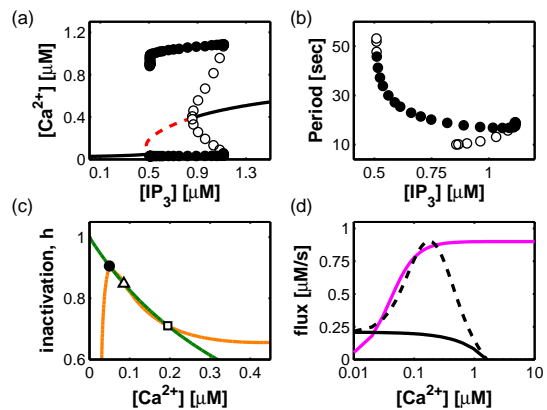}
\caption{\textbf{An excitable version of the Li-Rinzel model}.
\textit{(a)}~Bifurcation diagram and \textit{(b)}~period diagram of an excitable version of the LR model with K$_{3}=0.051$ $\mu$M. In this case, four bifurcations exist: a saddle-node and a saddle-node on invariant circle (SNIC) at [IP$_{3}$]$\simeq 0.479$ $\mu$M and [IP$_{3}$]$\simeq 0.526$ $\mu$M, and two subcritical Hopf bifurcations at [IP$_{3}$]$\simeq 0.51$ $\mu$M and [IP$_{3}$]$\simeq 0.857$ $\mu$M. \textit{(c)}~Between the two
saddle-node bifurcations nullclines intersect in three points which are a stable focus ($\bullet$) and an unstable node ($\square$) separated by a saddle ($\triangle$). \textit{(d)}~$J_{pump}$ (magenta curve) now intersects $J_{rel}$ at lower Ca$^{2+}$.}\label{fig2:ExLR}
\end{figure}
\indent We begin with the well studied and simpler AM dynamics that corresponds to higher values of $K_{3}$. As stated above, in this case a limit cycle emerges through a supercritical Hopf bifurcation where a single stable fixed point becomes unstable. In fig.~\ref{fig1:OrLR}c we show the nullclines of $h$ and $C$ when the fixed point is unstable. In this case the properties of the limit cycle can simply be understood from linear stability analysis of the unstable fixed point near the bifurcation. In fig.~\ref{fig1:OrLR}d we plot the calcium fluxes (for two different values of~$I$) as determined by setting $h$ to $h_{\infty} (C)$: these fluxes capture the fast time scale response of the system close to the fixed point, since the rate of receptor response is slower than that of the Ca$^{2+}$ concentration.~The fixed point occurs at {\em high} calcium, where the slope of the efflux $J_{rel}=J_{chan}+J_{leak}$ is negative. At values of Ca$^{2+}$ above the unstable fixed point, $J_{rel} < J_{pump}$, and there is a return flow in the ($h,C$) phase plane towards small values of $C$ and vice versa below the fixed point. The instability, then, gives rise to a dynamical flow that does not exhibit any strong positive amplification and instead is due to the slow dynamics. This fixes the frequency to be close to that of the $h$ time delay which does not vary much across the oscillatory regime. Conversely, the amplitude is relatively free to vary so that the system exhibits amplitude modulation in response to IP$_{3}$ variations.\\
%%
%%Figure 3 Schematic
\begin{figure}[bp!]
\includegraphics[width=0.42\textwidth]{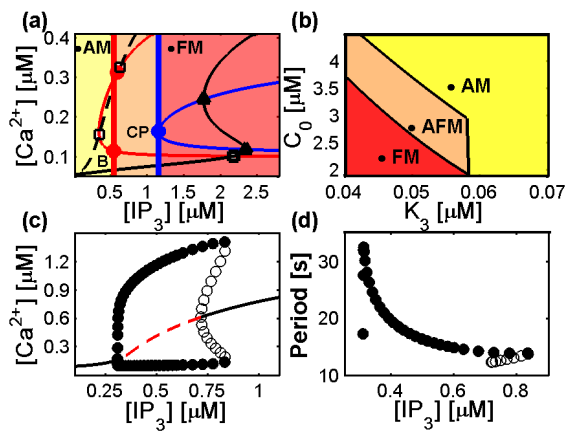}
\caption{\label{fig3:AFM} \textbf{AFM encoding} (colors online).
\textit{(a)}~Continuation of AM and FM bifurcations allows to identify a region comprised between a Bautin (B) and a cusp (CP) bifurcation where AFM-encoding could be found (($\square$) Hopf points, ($\triangle$) saddle-node points). \textit{(b)}~Modulation map illustrating the regime of existence of the three classes - AM, FM and AFM - of dynamical responses. \textit{(c)}~Bifurcation diagram and \textit{(d)}~period diagram for an AFM-encoding version of the LR model ($d_{5}=0.2$~$\mu$M, $C_{0}=4.0$~$\mu$M).}
\end{figure}
\indent
From the above analysis, it is clear that the key to get frequency modulation is to make the stable fixed point occur at \textit{low} calcium, on the rising part of the efflux curve. In our case this is accomplished by increasing the pumping rate by considering lower values of K$_{3}$, as shown in fig.~\ref{fig2:ExLR}. In such conditions in fact, when we consider the $h$ and $C$ nullclines (fig.~\ref{fig2:ExLR}c) we note that for proper $I$~values there is a stable fixed point at low Ca$^{2+}$ levels which is close to a saddle-point. Inspection of the characteristics of the Ca$^{2+}$ fluxes (shown in fig.~\ref{fig2:ExLR}d) reveals that at the saddle-point the slope of the characteristic of J$_{rel}$ is steeper than that of J$_{pump}$.~In this situation, a finite yet relatively small deviation away from the stable fixed point crosses the saddle-point separatrix and leads to a large excursion in the phase plane. For this reason, the Hopf bifurcation of the stable fixed point (which still occurs before $I_{sn}$ for the new set of parameters) must now be subcritical and different deviations (i.e. different values of $I$) lead to trajectories with similar amplitudes (fig.~\ref{fig2:ExLR}a), as this is determined by the global flow. At the same time the flow field in the vicinity of the stable fixed point and the saddle-point is very weak, hence the period of the excursions is very sensitive and can be effectively modulated by the IP$_{3}$ (fig.~\ref{fig2:ExLR}b). This dynamics shows ``Frequency Modulation".\\
\indent
The transition from AM to FM occurs via a characteristic codimension-2 bifurcation sequence which comprises the following steps: first, the lower supercritical Hopf point changes into a subcritical one via Bautin bifurcation; then, elsewhere in the parameter space, a cusp bifurcation generates the SNIC and saddle-node bifurcations which are responsible for variable-period oscillations (fig.~\ref{fig3:AFM}a). When the Hopf bifurcation is not yet strongly subcritical but we can nonetheless feel the influence of saddle-node coalescence, we can predict that the emerging oscillations would show significant variability both in amplitude and frequency (figs.~\ref{fig3:AFM}c,d,~\ref{fig3:Modulations}c). This is the previously mentioned AFM dynamics where IP$_{3}$ variations modulate both the amplitude and the frequency of the oscillations.\\
%Clearly, the extent of the parametric region comprised between the Bautin and cusp bifurcations defines the range of coexistence of AM and~FM~\cite{DePitta2007b}.\\
\indent
According to our analysis, AFM encoding can be found in several cases,  most typically for higher $C_{0}$ (fig.~\ref{fig3:AFM}b) or smaller $v_{3}$ values (table~\ref{table-table2}). In terms of the Ca$^{2+}$ fluxes (eqs.~\ref{eqs:Fluxes}), we note that such a choice of parameters counteracts the effect of a lower $K_{3}$ by increasing the distance between the characteristics, either by increasing $J_{rel}$ (through an increase of $C_{0}$) or by reducing $J_{pump}$ (by lowering $v_{3}$). It follows that the stable fixed point moves towards higher calcium levels and the amplitude modulation can coexist with frequency modulation.\\
\begin{table}[bp!]
\centering{
\begin{tabular}{|rr|rr|rr|rr|}
\hline
\multicolumn{2}{c}{{\bf Fixed}} & \multicolumn{2}{c}{{\bf $d_{5}=0.2$ $\mu$M}} & \multicolumn{2}{c}{{\bf $v_{2}=2\cdot10^{-3}$ s$^{-1}$}} & \multicolumn{2}{c}{{\bf $K_{3}=0.051$ $\mu$M}} \\

\multicolumn{2}{c}{{\bf Variable}}                  & \multicolumn{2}{c}{Range} & \multicolumn{2}{c}{Range} & \multicolumn{2}{c}{Range} \\
\hline
  {\bf $d_{5}$} &       [$\mu$M] &     --     &     --     &         0.107 &         0.161 &      0.085 &      0.127 \\
\hline
  {\bf $v_{2}$} &        [s$^{-1}$] &      0.175 &      0.262 &     --     &     --     &      0.178 &      0.266 \\
\hline
  {\bf $K_{3}$} &       [$\mu$M] &      0.134 &      0.201 &     --     &     --     &     --     &     --     \\
\hline
  {\bf $v_{3}$} &    [$\mu$Ms$^{-1}$] &      0.149 &      0.595 &      0.139 &      0.555 &      0.144 &      0.577 \\
\hline
  {\bf $C_{0}$} &       [$\mu$M] &      2.949 &      4.424 &      2.115 &      3.172 &      3.037 &      4.555 \\
\hline
\end{tabular}
} \caption[]{\footnotesize Parameter ranges for the coexistence of amplitude and frequency modulation of Ca$^{2+}$ response in Li-Rinzel model.} \label{table-table2}
\end{table}
%%
%%Figure 4
\begin{figure}[tbp!]
\includegraphics[width=0.4\textwidth]{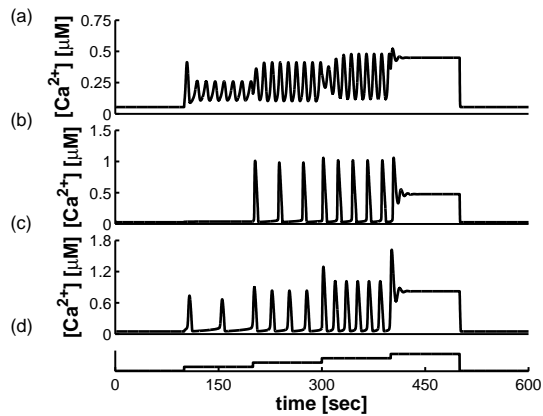}
\caption{\label{fig3:Modulations} \textbf{Different types of excitability}. Proper tuning of parameters allows for different Ca$^{2+}$ responses for a generic IP$_{3}$ stimulus (depicted in~\textit{(d)}). \textit{(a)}~The original LR parameters (table~\ref{table-table1}) provide amplitude variability of oscillations that occur at almost fixed frequency. \textit{(b)}~A higher SERCA pump Ca$^{2+}$ affinity ($K_{3}=0.051$~$\mu$M) is responsible for oscillations with variable frequency but nearly constant amplitude. Despite the different nature of bifurcations underlying these two cases, both AM and FM features can also be found for example in presence of higher cell-averaged total free Ca$^{2+}$ levels as shown in \textit{(c)} for $C_{0}=4$~$\mu$M (with $K_{3}=0.051$~$\mu$M). Stimulus: \textit{(a)}~$\Delta$IP$_{3}=0.4$, $0.1$, $0.1$, $0.5$~$\mu$M; \textit{(b)}~$\Delta$IP$_{3}=0.4$, $0.2$, $0.2$, $0.5$~$\mu$M; \textit{(c)} $\Delta$IP$_{3}=0.17$, $0.03$, $0.09$, $0.5$~$\mu$M.}
\end{figure}
\indent
In summary, we showed the existence of three distinct classes of information encoding modes in the response of Ca$^{2+}$ to IP$_{3}$ variations - AM, FM and AFM.~Two of these classes, AM and FM, were previously demonstrated in other minimal models \citep{FalckeRev2004} whereas AFM has been explained so far only by extended models which included either diffusive terms \cite{FalckeRev2004} or complex feedback loops \citep{PolitiHofer2006,KummerBaier2000,MarhlBioelectrochemBioenerg1998}. We found that AFM dynamics could also be reproduced by a minimal model for some particular range of the model parameters. These findings hint that by activating intracellular mechanisms that control the values of physiological parameters that correspond to the model parameters C$_{0}$ or v$_{2}$, K$_{3}$ and v$_{3}$ for SERCA pumps or d$_{5}$ for IP$_{3}$Rs~\citep{Toescu1995}, the type of information encoding can be regulated.\\
\indent
We note that for the different modes of Ca$^{2+}$ response to have an information encoding role, a corresponding decoding mechanism must exist. The existence of decoding mechanisms of AM and FM have been proposed in \citep{LarsenKummer2004} based on model studies of the cooperative binding of Ca$^{2+}$ to generic effector enzymes. In principle, the same mechanism can also decode information embedded in Ca$^{2+}$ that corresponds to AFM.\\
\indent
In the context of communication theory, cellular Ca$^{2+}$ signalling can be regarded as a \textit{bifurcation-based} encoding system: the baseline IP$_{3}$ level I$_{0}$ is set to be sufficiently close to a bifurcation point so that the variations in IP$_{3}$ caused by external signals regularly cross that point. This mechanism can be in principle realized in new kind of electronic systems. In AM, Ca$^{2+}$ peaks encode the information; in FM, variations in the IP$_{3}$ will trigger bursts of Ca$^{2+}$ spikes (fig.~\ref{fig3:Modulations}) with information encoded in the inter-spike intervals. In the mixed AFM mode, both features carry information which can be separately decoded by different downstream effectors with different Ca$^{2+}$ responses. This is particularly suitable for those systems that require special constraints in coordination of informational input from multiple channels, for which recently, mixed AM and FM modulation (MM) has been receiving much attention.\\
\indent
The above can be very relevant for the case of astrocytes regulation of synaptic information transfer. Astrocytes respond to synaptic activity through their intracellular Ca$^{2+}$ dynamics which in turn feeds back to neurons by triggering release of gliotransmitter. AFM encoding could have deep consequences \cite{Carmignoto2000}. We might expect that the short-time scale effectors (mostly sensitive to the number of pulses) are involved in feedback to the local synapse whereas the long-time scale ones (which integrate the total signal) coordinate information with other astrocytes via intercellular signalling.\\
\indent The authors thank~V.~Parpura,~G.~Carmignoto, M.~Zonta,
B.~Ermentrout, B.~Sautois and N.~Raichman for insightful
conversations.~V.V. was supported by ICAM Travel Award.~This
research has been supported by the NSF-sponsored Center for
Theoretical Biological Physics (grant~nos.~PHY-0216576 and
PHY-0225630) and by the Tauber Fund at Tel-Aviv~University.
\bibliography{depitta07}
\end{document}